\documentclass[%
 reprint,
superscriptaddress,
 amsmath,amssymb,
 aps,
 pra,
]{revtex4-2}

\usepackage{dcolumn}
\usepackage{bm}
\usepackage{mathtools} 
\usepackage{color}  

\definecolor{mscolor}{rgb}{0,0.5,0.5}

\definecolor{tgcolor}{rgb}{0.5,0,0.5}

{}
\definecolor{phcolor}{rgb}{0.5,0,0.5}
\newcommand\T{\rule{0pt}{4ex}}       
\newcommand\B{\rule[-2ex]{0pt}{0pt}} 

\begin{document}



\title{A simple, passive design for large optical trap arrays for single atoms}

\title{A simple, passive design for large optical trap arrays for single atoms}

\author{P. Huft}%
 \email{huft@wisc.edu}
\affiliation{%
Department of Physics, University of Wisconsin-Madison, 
 1150 University Avenue,
 Madison, WI 53706
}
\author{Y. Song}
\affiliation{%
Department of Physics, University of Wisconsin-Madison, 
 1150 University Avenue, 
 Madison, WI 53706
}

\author{T. M. Graham}
\affiliation{%
Department of Physics, University of Wisconsin-Madison, 
 1150 University Avenue, 
 Madison, WI 53706
}

\author{K. Jooya}
\affiliation{%
Department of Physics, University of Wisconsin-Madison, 
 1150 University Avenue, 
 Madison, WI 53706
}

\author{S. Deshpande}
\author{C. Fang}
\author{M. Kats}
\affiliation{%
Department of Electrical and Computer Engineering, University of Wisconsin-Madison, 
 1415 Engineering Drive, 
 Madison, WI 53706
}

\author{M. Saffman}
\affiliation{%
Department of Physics, University of Wisconsin-Madison, 
 1150 University Avenue, 
 Madison, WI 53706
}
\affiliation{ColdQuanta, Inc.,  111 N. Fairchild St., Madison, WI 53703}

\date{\today}

\begin{abstract}
We present an approach for trapping cold atoms in a 2D optical trap array
generated with a novel $4f$ filtering scheme and custom transmission
mask without any active device. The approach can be used to
generate arrays of bright or dark traps, or both simultaneously with a single wavelength for forming two-species traps. We demonstrate the design by creating a 2D array of 1225 dark trap sites, where single Cs atoms are loaded into regions of near-zero intensity in an approximately
Gaussian profile trap. Moreover, we demonstrate a simple solution
to the problem of out-of-focus trapped atoms, which occurs due to
the Talbot effect in periodic optical lattices. Using a high power yet low cost  spectrally and spatially broadband
laser, out-of-focus interference is mitigated, leading to near perfect
removal of Talbot plane traps.
\end{abstract}

\maketitle

\section{Introduction}
Optical trap arrays are a key ingredient in neutral atom based quantum technologies, including quantum computing, quantum simulation, and quantum sensing, due to their stability and versatility \cite{Kaufman2021,Morgado2021}. This is the result of advances over the past two decades in creating low-entropy arrays of single atoms, which have made optical trap arrays ubiquitous in quantum science. However, the optical setups for creating these traps are often complicated, space-consuming, and expensive, requiring active electro-optical devices such as liquid crystal spatial light modulators (SLMs)\cite{DKim2019}, acousto-optic deflectors (AODs)\cite{Graham2022}, and digital micromirror devices (DMDs)\cite{YWang2020}. In response to this experimental overhead, we propose a simple method of creating optical trap arrays using only passive components, consisting of a mask with a custom transmission pattern and a $4f$ imaging setup with a Fourier plane iris for spatial filtering.

Two major advantages exist for optical traps created with passive rather than active components: 1) passive components are free from noise associated with active devices, such as intensity flicker which can lead to short trap lifetimes \cite{Stuart2018}, and 2) they have the capability to handle high optical powers which enables scaling to very large trap arrays. The approach we demonstrate uses a passive amplitude mask, which has some advantages over a passive hologram. Specifically, an amplitude mask can be used with a broad range of wavelengths, and can also be used with incoherent light, which we show can be used for mitigating the Talbot effect for periodic trap arrays. We show how the same basic working principle can be used to create both bright and dark traps, where atoms are trapped in regions of high and low intensity, respectively. Moreover, this approach can be used to create a dual-grid of both dark and bright traps for confining two different atomic species\cite{Singh2022} using only one passive optical mask and a single trapping wavelength. We show as a proof of principle the creation of a two-dimensional 1225-site  array which is used for trapping single Cs atoms in blue-detuned dark traps. Several other recent experiments have demonstrated large atom arrays \cite{YWang2020,Scholl2021,Ebadi2021}. However, all of these relied on active devices. A microlens array can be used to create an array of red-detuned bright traps without requiring any active devices\cite{deMello2019}, but cannot be easily extended to the dark and bright-dark arrays described here.  

The paper is organized as follows. In sec. \ref{sec:theory}, the working principle is discussed for creating an array of bright red-detuned traps, dark blue-detuned traps, or a combination of the two. In sec. \ref{sec:exp}, we discuss an experimental demonstration of trapping single Cs atoms in a dark trap array, and show that the Talbot effect can be mitigated by using incoherent trap light. 

\section{Working Principle} \label{sec:theory}

\subsection{Bright Trap Array}
\begin{figure*}
    \includegraphics[width=\textwidth]{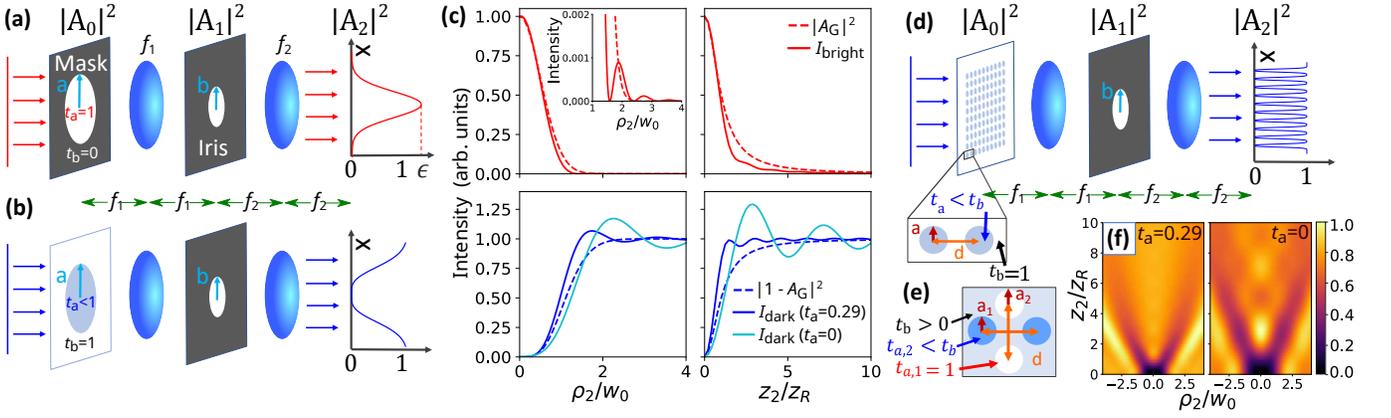}
    \caption{Working principle for creating approximately Gaussian (aG) field profiles using a $4f$ filtering scheme. \textbf{(a)} The design for producing a bright trap from an approximately Gaussian beam. An input plane wave is apertured by an opaque mask with a transmitting hole of radius $a$ to create a top-hat profile. A spatial filter or iris of radius $b$ in the Fourier plane transmits only the central lobe of the Airy disk. After transforming through lens $f_2$, the output field, $A_2$, is an approximately Gaussian (aG) beam. \textbf{(b)} Dark traps can be produced with a mask which is fully transmitting with a partially transmitting aperture, for which the output is the aG field subtracted from a plane wave. \textbf{(c)} Intensity profiles of bright (red) and dark (blue) traps based on aG beams (solid lines) compared to their Gaussian (dashed lines) counterparts plotted in terms of the Gaussian waist $w_0$ and Rayleigh range $z_R=\pi w_0^2/\lambda$ for the Gaussian bright trap. The bright traps are normalized to their respective peak focal plane intensities and the dark traps are normalized to the asymptotic intensity at large $\rho_2,z_2$. The inset shows a zoomed in view of the higher order deviation between the aG and Gaussian beam. The axial ($\rho_2=0$) and radial ($z_2=0$) profiles for the bright and dark aG trap are computed numerically using Fresnel diffraction theory. The pairs of dark and light solid blue curves have $t_a=0.287$, $t_a=0$, respectively. For all aG beam curves,  $a=100 ~\mathrm{ \mu m}$ and $b=f_1 x_1^{(1)}/a k$, except for the light blue ($t_a=0$) curve, which has $b=f_1 x_1^{(0)}/a k$. The other simulation parameters are $\lambda=825$ nm, and $w_0=0.974(f_2/f_1)a$ and $w_0=0.943(f_2/f_1)a$ for the bright and dark Gaussian beams, respectively. \textbf{(d)} Extension of the $4f$ filtering scheme to produce a 2D grid of dark traps, using a mask with a grid of apertures with spatial period $d$. \textbf{(e)} A dual grid of bright and dark traps with a single wavelength can be made with a mask which has finite background transmission $t_b$ populated a grid of fully transmitting ($t_{a,1}=1$) apertures and a dual grid of apertures with transmission $t_{a,2} < t_b$. \textbf{(f)} Dark trap profiles in the $\rho z$ plane corresponding to the blue and light blue solid curves in (c), normalized to their respective peak intensities.}
    \label{fig:working_principle}
\end{figure*}

Here we discuss the case of creating an array of bright optical dipole traps, in which atoms are trapped in regions of maximum intensity for light which is far-detuned red of an atomic transition. Consider a plane wave, incident on an opaque mask with a fully transmitting aperture of radius $a$ (Fig. \ref{fig:working_principle}). If the illuminated mask is placed in the front focal plane of a lens $f_1$, the field in the back focal plane is the familiar Airy disk, which is the Fourier transform of the top hat profile of the field just after the mask. Because the Fourier plane field gives the spatial frequency spectrum of the front focal plane field, we can reason that filtering out higher spatial frequencies from the Airy disk will have the effect of creating a low-passed top hat beam after a second lens transformation.  
Placing an iris of radius $b$ in the Fourier plane to filter the Airy disk and transform through lens $f_2$, the output field in the back focal plane found from Fresnel diffraction theory is 
\begin{equation}
A_{2}\left(\rho_{2}\right)=-A_{0} \frac{a k}{f_{2}} \int_{0}^{b} d \rho_{1} J_{0}\left(\frac{\rho_{2} k}{f_{2}} \rho_{1}\right) J_{1}\left(\frac{a k}{f_{1}} \rho_{1}\right).
\label{eq:a2_field}
\end{equation}
 The finite Bessel integral above can be expressed as a power series in $b$ using 
\begin{eqnarray}
     &&\int_{0}^{b} d z J_{0}(c z) J_{1}(d z) \nonumber\\
     &=& 
  \sum_{j=0}^{\infty} \frac{(-1)^{j}}{j !(j+1) !(2 j+2)} \nonumber\\
  &\times&  { }_{2} F_{1}\left(-j,-1-j ; 1 ; c^{2} / d^{2}\right) b^{2+2 j}(d / 2)^{1+2 j}\nonumber
\end{eqnarray}
where $_2F_1$ is the hypergeometric function \cite{pochernyaev1995}. Taking $f_1 = f_2 = f$ and setting $b=\frac{f}{a k} x_1^{(1)}$, where $x_1^{(1)}=3.8317$ is the first zero of $J_1$, allows further simplification. This choice for $b$ corresponds to filtering off the lobes beyond the central bright spot, constituting a power loss of only 16$\%$. With these choices, we can then express the intensity in the output plane $I_2$, normalized to the input intensity $I_0$, as a power series in $\rho_2/a$:
\begin{equation}\label{eq:Iag_radial}
    \frac{I_{2}\left(\rho_{2},z_2=0\right)}{I_{0}}=1.97-4.15\left(\frac{\rho_{2}}{a}\right)^{2}+3.92\left(\frac{\rho_{2}}{a}\right)^{4}-\ldots
\end{equation}
To compute the on-axis expansion of the intensity, we modify the integral in eq. (\ref{eq:a2_field}) by setting $\rho_2=0$ and including the quadratic phase factor $\text{exp}\big(-i\rho_1^2 z_2/2 f_2^2\big)$ in the integrand, where $z_2$ is the axial deviation from the back focal plane of the second lens. Both the radial and axial expansions, renormalized to have peak value of 1, are
\begin{equation}\label{eq:Iag_bright}
\begin{aligned}
\frac{I_{2}\left(\rho_{2},z_2=0\right)}{I_2(0,0)}&=1-2.11\left(\frac{\rho_{2}}{a}\right)^{2}+1.99\left(\frac{\rho_{2}}{a}\right)^{4}-\ldots\\
\frac{I_2(\rho_2=0,z_2)}{I_2(0,0)}&= 1 - 2.60 \frac{z_2^2}{a^4 k^2} + 3.28\frac{z_2^4}{a^8 k^4} - ...\\
\end{aligned}
\end{equation}
We will refer to this intensity profile as an approximately Gaussian (aG) beam. Equating the radial profile to a Gaussian intensity profile $|A_G(\rho_2)|^2 = \exp(-2 \rho_2^2/w_0^2))$ at quadratic order, we find that the aG beam is, to a very good approximation, a Gaussian beam with $1/e^2$ waist $w_0=0.974 (f_2/f_1)a$  (Fig. \ref{fig:working_principle}). 
 
It is useful to recast the coefficients of the quadratic terms for the radial and axial expansions of $I_2$ in terms of Gaussian beam parameters. This is useful for seeing how the trap confinement compares to a pure Gaussian optical trap at lower-order in $\rho,z$, and for ease in modifying common expressions pertaining to Gaussian optical traps, such as trap frequencies (see eq. (\ref{eq:harmonic_freqs})). We already showed that the radial expansion can be fit to a Gaussian with $w_0=0.974 a$, so we can then  equate the quadratic coefficient of the axial expansion in eq. (\ref{eq:Iag_bright}) to a numerical factor times $1/z_R^2$, where $z_R=\pi w_0^2/\lambda$ is the familiar Rayleigh range of a Gaussian beam.
The radial and axial expansions, expressed in terms of Gaussian parameters up to quadratic order and normalized to the peak intensity $I_2(0,0)$ are
\begin{equation}\label{eq:Iag_bright2}
    \begin{aligned}
    \frac{I_2(\rho_2,z_2=0)}{I_2(0,0)} &= 1 - 2 \left(\frac{\rho_2}{w_0}\right)^2 + 1.79 \left(\frac{\rho_2}{w_0}\right)^4 - ... \\
    \frac{I_2(\rho_2=0,z_2)}{I_2(0,0)} &= 1 - 0.59 \left(\frac{z_2}{z_R}\right)^2 + 0.17 \left(\frac{z_2}{z_R}\right)^4 - ...
    \end{aligned}
\end{equation}
We see that the radial confinement of the bright aG beam is the same as that of the best-fit Gaussian beam, whereas the axial confinement is looser by around 30$\%$.

This $4f$ filtering approach to creating an aG beam is readily extended to create an array of aG beams. For example, a 2D grid of $n \times n$ aG beams can be generated by making the input mask a 2D square grid of $n \times n$ apertures. The field transmitted through each aperture will be an aG pattern as derived above, but will appear at position $-\rho_{ij}$ in the output plane, where $\rho_{ij}$ is the input position of the corresponding aperture. This is shown for a 2D square array in Fig. \ref{fig:working_principle}. Provided the spatial period $d$ of apertures on the mask satisfies $d \gtrsim 3a$, the interference between adjacent beams will be negligible. The resulting array of traps can then be re-imaged with relay optics to create small  traps suitable for single atom trapping. 

The efficiency of the trap creation is defined as the ratio of peak intensity of the trap in the output plane to the intensity of the input plane wave incident on the mask. This is given by $=I_{t}/I_{d}$ where $I_{d} = P / d^{2}$ is the intensity of an input $d \times d$ unit cell, and $I_{t} = I_2(\rho_2=0,z2=0)$, which is simply  eq. (\ref{eq:Iag_radial}) evaluated at $\rho_2=0,z_2=0$. Hence, independent of the aperture radius $a$, the efficiency is given by
\begin{equation}\label{eq:brighteff}
    \epsilon = \frac{I_{t}}{I_{d}} = \frac{I_{2}\left(0,0\right)}{I_{0}}=1.97
\end{equation}
The meaning of an efficiency greater than unity is that the input light has been redistributed to form a more localized region of intensity with a profile suitable for trapping. Note that this value of efficiency assumes equal focal lengths, $f_1=f_2$. For unequal focal lengths, $\epsilon$ scales as the magnification factor for intensity, $(f_1/f_2)^2$. 
The fraction of optical power transmitted through the $4f$ filtering system is equal to the fraction of power transmitted through the mask, times the fraction transmitted through the Fourier filter. This gives $ P_{\rm out}/P_{\rm in}= 0.84(\pi a^2/d^2)$. For example, with $d=3a$, $P_{\rm out}/P_{\rm in}=0.29$. As we will see, the design for creating dark traps is  more efficient in terms of power transmission.

We note that a single array mask can be used with different laser wavelengths $\lambda$ simply by adjusting the Fourier filter radius $b$.

It is also possible to create tighter trap profiles than those described above, at the expense of using a  more complicated amplitude or phase mask in the Fourier plane\cite{Beguin2020}. This is discussed in Appendix \ref{sec:tighter}.

\subsection{Dark Trap Array}
The scheme presented above can be modified to produce dark traps, where atoms are trapped in regions of zero intensity for blue-detuned trap light. Dark traps are often preferable over bright traps for a number of reasons. For example, 1) the trapped atoms are insensitive to laser power fluctuations 2) the trap light can be kept on during experiment sequences involving laser excitation, which may be untenable in bright traps due to large AC Stark shifts, and 3) Rydberg states, for which the AC Stark shift is always positive, can also be trapped\cite{SZhang2011}.

It is possible to form an aG dark trap, in which an atom can be trapped at the intensity minimum, and axial confinement is provided by the diffraction of the field out of focus. This can be done with a mask which is somewhat complementary to that above, having partially transmitting apertures on a fully transmitting background. Denoting the aperture and background transmission amplitudes as $t_a$ and $t_b$, respectively, the field after transforming to the Fourier plane is
\begin{widetext}
\begin{eqnarray}
          A_{1}\left(\rho_{1}\right) &=&-i \frac{A_{0} k}{f_{1}}\bigg[t_{a} \int_{0}^{a} d \rho_{0}\, \rho_{0} J_{0}\left(\frac{k \rho_{0} \rho_{1}}{f_{1}}\right) +t_{b} \int_{a}^{\infty} d \rho_{0}\, \rho_{0} J_{0}\left(\frac{k \rho_{0} \rho_{1}}{f_{1}}\right)\bigg]\nonumber
         \\
  &=&-i \frac{A_{0} k}{f_{1}}\bigg[\left(t_{a}-t_{b}\right) \int_{0}^{a} d \rho_{0}\, \rho_{0} J_{0}\left(\frac{k \rho_{0} \rho_{1}}{f_{1}}\right) 
            +t_{b} \int_{0}^{\infty} d \rho_{0}\, \rho_{0} J_{0}\left(\frac{k \rho_{0} \rho_{1}}{f_{1}}\right)\bigg].
\end{eqnarray}
\end{widetext}
The second term in the square brackets, after filtering in the Fourier plane and a second lens transformation, gives a plane wave with amplitude $-t_{b} A_{0}$. The first term gives the same result derived for the field in eq. (\ref{eq:a2_field}) multiplied by $t_{a}-t_{b}$, such that the total field is a plane wave minus an aG near-Gaussian profile. For $f_{1}=f_{2}=f$, the on-axis field in focus is given by
\begin{equation}\label{eq:darkcenter}
    A_{2}(0)=-A_{0}\left[t_{b}+\left(t_{a}-t_{b}\right)\left[1-J_{0}\left(\frac{k a b}{f}\right)\right]\right]
\end{equation}
This leads to the condition for the field to be zero
\begin{equation}\label{eq:darkcondition}
    t_{a}=-t_{b} \frac{J_{0}\big(\frac{k a b}{f}\big)}{1-J_{0}\big(\frac{k a b}{f}\big)}.
\end{equation}
Choosing $b=\frac{f}{a k} x_1^{(1)}$ gives $t_a=0.287t_b$.
For a fully transmitting background, $\left| t_b \right|=1$, which implies an aperture transmission $T_a = \left| t_a \right| ^2=0.082$ to make a dark near-Gaussian trap with a zero intensity minimum. It should be clear that $f_1$ need not equal $f_2$ for this result for the aperture transmission amplitude to hold true. The sensitivity of this zero intensity condition to the iris radius and relative phase between $t_a$ and $t_b$ is discussed in Appendix \ref{sec:sensitivity}. 

The radial and axial expansions of the dark aG trap are  
\begin{equation} \label{eq:Iag_dark}
    \begin{aligned}
    \frac{I_2(\rho_2,z_2=0)}{I_0} &= -1.04\times10^{-6} \left(\frac{\rho_2}{w_0}\right)^2 + 1.11 \left(\frac{\rho_2}{a}\right)^4 - ... \\
    \frac{I_2(\rho_2=0,z_2)}{I_0} &= \frac{4.44}{a^4 k^2}z_2^2 - \frac{8.04 }{a^8 k^4}z_2^4 + ...
    \end{aligned}
\end{equation}
 The quadratic term in the radial expansion is very close to zero, and will be dropped going forward. Following the analogy between the Gaussian and aG fields, this trap profile is similar to $|1 - A_G|^2$, for which the first non-vanishing term in the radial expansion at $z_2$=0 is quartic. Matching the quartic term of eq. (\ref{eq:Iag_dark}) to that of the Gaussian-based equivalent, we find $w_0=0.943a(f_2/f_1)$, where the focal lengths have been left arbitrary. Again, we can compare the expansion with a Gaussian-based trap by recasting the coefficients in terms of Gaussian parameters as was done for the bright trap:
\begin{equation} \label{eq:Iag_dark2}
    \begin{aligned}
    \frac{I_2(\rho_2,z_2=0)}{I_0} &= \left(\frac{\rho_2}{w_0}\right)^4 - ... \\
    \frac{I_2(\rho_2=0,z_2)}{I_0} &= 1.01\left(\frac{z_2}{z_R}\right)^2 - 0.33 \left(\frac{z_2}{z_R}\right)^4 + ...
    \end{aligned}
\end{equation}
The trap profiles from eqs. (\ref{eq:Iag_bright2}) and (\ref{eq:Iag_dark2}) are plotted in Fig. \ref{fig:working_principle} alongside their Gaussian counterparts.

The dark aG trap radial profile in the focal plane is nearly quartic. One consequence is that the distribution of atoms will therefore be different than for a harmonic trap, as discussed in section \ref{sub:confine}. A trapping potential which is harmonic to lowest order may be desirable in some cases, for example to allow for the implementation of sideband cooling \cite{Wineland1979}. For particular values of finite $t_a$, the traps generated with this design can be made harmonic by imposing a finite phase difference $\phi_{ab}$ between the transmitting mask background and choosing a suitable iris radius $b$. This is discussed further in Appendix \ref{sub:profiles}.

An attractive modification of the dark aG trap is to use a mask which has $t_a=0$, corresponding to opaque disks on the fully transmitting background (the complement of the bright trap mask), which may be easier to fabricate reliably compared to the version requiring a specific finite value for $t_a$. This could be implemented with either a passive optical element or or an active amplitude spatial light modulator such as a DMD. From the condition for a dark trap center given in eq. (\ref{eq:darkcondition}), we find that the iris radius should be set to $b_n=(f/ka)x_n^{(0)}$, where $x_n^{(0)}$ is the $n^{\rm th}$ zero of Bessel $J_0$ and $n>0$. In the limit of large $n$, the trap radial profile   approaches a square well of radius $a$, which is simply the re-imaged mask aperture with no Fourier filtering. For iris radius $b_1$ and $a=w_0(f1/f2)/0.943$, the radial and axial trap profiles in terms of Gaussian beam parameters are given by
\begin{equation} \label{eq:Iag_dark3}
    \begin{aligned}
    \frac{I_2(\rho_2,z_2=0)}{I_0} &= 0.31 \left(\frac{\rho_2}{w_0}\right)^4 - 0.12 \left(\frac{\rho_2}{w_0}\right)^6... \\
    \frac{I_2(\rho_2=0,z_2)}{I_0} &= 0.31 \left(\frac{z_2}{z_R}\right)^2 - 0.03 \left(\frac{z_2}{z_R}\right)^4 + ...
    \end{aligned}
\end{equation}
and  shown in Fig. \ref{fig:working_principle}. 

The efficiency of the dark aG trap for all variations considered is given approximately by
\begin{equation}
\epsilon=\frac{I_{t}}{I_{d}}=\frac{I_{d}}{I_{d}}=1
\end{equation}
which follows from the fact that the input plane wave is fully transmitted through the Fourier filter. For the dark traps with $t_a=0.287$ and $t_a=0$ shown in Fig. \ref{fig:working_principle}, the efficiency is about 1.1 and 1.2, respectively, due to diffraction effects. For an array of dark traps, this efficiency is valid when interference between neighboring traps is negligible with $d\gtrsim6a$. The efficiency of the dark trap variants is lower than for the bright trap, but compares favorably with dark traps created with a Gaussian beam array using diffractive optical elements which has $\epsilon \leq 0.51$ \cite{Piotrowicz2013} or a line array which
has $\epsilon \leq 0.97$. \cite{Saffmanlines}.

The fractional power transmission through the $4f$ filtering setup for the dark trap array is more favorable than that of the bright trap, which has a mask with an opaque background. For a mask with arbitrary background and aperture transmissions, we have $P_{\rm out}/P_{\rm in} = \eta(|t_b|^2(d^2 - \pi a^2) + |t_a|^2 \pi a^2)/d^2$ where $\eta$ is the fractional transmission of the Fourier filter, which depends on the optimal filter radius for the choice of $t_a$. For $t_a=0$,  $\eta=0.73$, found by integrating the Airy disk up to radius $b_1$ as defined above. For a mask with either $|t_a|=0$ or  $|t_a|=0.287$ and $d=3a$, $P_{\rm out}/P_{\rm in}\approx 0.50$.

\begin{table*}[!th]
    \centering
    \begin{tabular}{|c|c|c|c|}
    \hline
    Parameter & Gaussian & Bright aG & Dark aG \\
    \hline \T
    $I(\rho,0)/I(0,0)$ & $1 - 2\big(\frac{\rho}{w_0}\big)^2 + 2 \big(\frac{\rho}{w_0}\big)^4 - ...$  & $1 - 2\big(\frac{\rho}{w_0}\big)^2 + 1.79 \big(\frac{\rho}{w_0}\big)^4 - ...$ & $ \big(\frac{\rho}{w_0}\big)^4 - ...$ \B \\
    \hline
    \T    
    $I(0,z)/I(0,0)$ & $1 - \big(\frac{z}{z_R}\big)^2 + \big(\frac{z}{z_R}\big)^4 - ... $ & $1 - 0.585 \big(\frac{z}{z_R}\big)^2 + 0.166 \big(\frac{z}{z_R}\big)^4 - ...$ & $ 1.01\big(\frac{z}{z_R}\big)^2 - 0.330 \big(\frac{z}{z_R}\big)^4 + ...$ \B \\
    \hline \T
    $\omega_{\rho}$ & $\frac{2}{w_0}\sqrt{\frac{U_0}{m}}$ & $\frac{2}{w_0}\sqrt{\frac{U_0}{m}}$ & ill-defined \B \\
    \hline \T
    $\omega_z$ & $\frac{1}{z_R}\sqrt{\frac{U_0}{m}}$ & $\frac{1}{1.307z_R}\sqrt{\frac{U_0}{m}}$ & $\frac{1}{0.997 z_R}\sqrt{\frac{U_0}{m}}$ \B \\
    \hline \T
    $\sigma_{\rho}$ & $w_0\big(\frac{k_B T}{2U_0}\big)^{1/2}=0.22~\mu$m & $w_0\big(\frac{k_B T}{2U_0}\big)^{1/2}=0.22~\mu$m & $w_0\big(\frac{2}{3} \frac{k_B T}{U_0}
    \big)^{1/4}= 0.28~\mu$m \B \\
    \hline \T
    $\sigma_z$ & $z_R\big(\frac{k_B T}{2U_0}\big)^{1/2}=0.87~\mu$m & $1.307 z_R\big(\frac{k_B T}{2U_0}\big)^{1/2}=1.14~\mu$m & $0.997 z_R\big(\frac{k_B T}{2U_0}\big)^{1/2}=0.87~\mu$m \B \\
    \hline
    \end{tabular}
    \caption{Comparison of normalized intensity profiles $I$, atom distribution standard deviations $\sigma$, and vibrational frequencies $\omega$ for the bright and dark aG ($t_a=0.287$) trap potentials, and a standard Gaussian bright trap. For ease of comparison, the expressions have all been cast in terms of the best-fit Gaussian waist $w_0$ and corresponding $z_R$. Numerical values for the standard deviations are given using $\lambda = 808$ nm and $w_0=1~\mu$m, and a ratio of atom kinetic energy to trap potential $k_B T/U_0 = 1/10$.}
    \label{tab:params}
\end{table*} 
\subsection{Combined Bright and Dark Trap Array}

An   interesting application of this technique is a grid of both bright and dark traps for trapping two atomic species with a single trapping wavelength. There has been interest in such two-species trap arrays as they have potential applications for error-corrected quantum computing \cite{Beterov2015}. While a large array of Rb and Cs atoms was recently demonstrated \cite{Singh2022}, the proposed approach requires only one trapping wavelength and a single passive optical element to create the intensity pattern for both traps, which simplifies the complexity  of the experimental setup. Such a trap can be made utilizing transitions in two species which have dynamic polarizabilities of comparable magnitude but opposite sign for the chosen wavelength.

Consider a mask with a background of transmission $t_b$, populated with a grid of fully-transmitting apertures for  bright traps and a dual grid of apertures with transmission amplitude $t_a = 0.287 t_b$ for forming dark traps, with $\left|t_a\right| < \left|t_b\right|$ (Fig. \ref{fig:working_principle}). The peak intensity of the bright trap occurs at $(\rho_2,z_2)=(0,0)$, and that of the dark trap occurs off-axis for large $\rho_2$, where the intensity is simply that of the plane wave transmitted through the mask. Assuming $f_1 = f_2$ for simplicity, and using eq. (\ref{eq:darkcenter}), the peak intensity of the output bright traps relative to the background intensity, and that of the dark traps are given by

\begin{widetext}
\begin{eqnarray}
    I_{\text {bright}}(\rho_2=0,z_2=0) &=
    &\bigg|\left(t_{b}-1\right)\left[1-J_{0}\left(x_1^{(1)}\right)\right]  - t_{b}\bigg|^{2} -\left| t_{b} \right| ^{2}\\
    I_{\rm dark}(\rho_2\rightarrow\infty,z_2=0) &=& \left|t_b \right|^2
\end{eqnarray}
\end{widetext}

where $I_{\rm bright}(\rho_2=0,z_2=0)$ and $I_{\rm dark}(\rho_2\rightarrow\infty,z_2=0)$ are found from eqs. (\ref{eq:Iag_radial}) and (\ref{eq:Iag_dark}), respectively. These relative intensities are equal for $t_b\approx0.77$.

The condition for one species to be trapped in the bright spots and one in dark traps, with equal trap depths, is  $\left|\alpha_b I_{\rm bright}\right| = \left|\alpha_d I_{\rm dark}\right|$. Choosing $\lambda=810$ nm, we can create bright traps for Rb and dark traps for Cs. The dynamic ground state polarizability at 810 nm for Rb is 
$\alpha_b = 847~ \textrm{\AA}^{3}$ and for Cs is $\alpha_d = -433~ \textrm{\AA}^{3}$, from which we obtain $t_b= 0.86$ to have equal trap depths for Rb and Cs. For the case of using fully opaque dark apertures, i.e. $t_a=0$, the dark trap efficiency is about 1.2 and we get $t_b=0.84$ for the bright and dark mask. Note that in this case, the dark apertures should have radius $a_{\textrm{dark}} = a_{\textrm{bright}} (x_1^{(0)}/x_1^{(1)})$ so that the same Fourier filter radius is optimal for both the bright and dark traps.

\subsection{Atom Confinement}\label{sub:confine}

The figures of merit for trapping a particle, such as an atom or molecule, are the depth of the trapping potential and the spatial confinement of the particle. Assuming the particle to be in a low energy motional state, we can approximate the trap as a harmonic potential by keeping only up to the quadratic terms in spatial coordinates. For a bright trap, where atoms are trapped at the peak intensity, we have $U=U_{0}\left(1-\epsilon_{\perp} \rho^{2}-\epsilon_{\|} z^{2}+\ldots\right)$, where $\rho$ and $z$ are the radial and axial coordinates of the particle, respectively, and $z$ is along the axis of optical propagation of the trap light. For a dark trap, where atoms are trapped at the minimum intensity, $U=U_{0}\left(\epsilon_{\perp} \rho^{2}-\epsilon_{\|} z^{2}+\ldots\right)$. Because the equations of motion are not affected by constant terms in the potential, the equations for trap frequencies and confinement that follow are valid for either bright or dark potentials.

We can obtain the spread in a trapped particle's position by using the Virial theorem to relate the potential and kinetic energy of the trapped particle. For a particle of temperature $T$, the standard deviations of the particle position are given by
\begin{equation}
\begin{aligned}
2 U_{0} \epsilon_{\perp}\left\langle\rho^{2}\right\rangle &=2 k_{B} T \\
2 U_{0} \epsilon_{\|}\left\langle z^{2}\right\rangle &=k_{B} T
\end{aligned}
\end{equation}
with $k_B$ the Boltzmann constant. The standard deviations of the particle position are therefore
\begin{equation}
\begin{gathered}
\sigma_{\rho}=\sqrt{\left\langle\rho^{2}\right\rangle}=\frac{1}{\epsilon_{\perp}^{1 / 2}}\left(\frac{k_{\mathrm{B}} T}{U_{0}}\right)^{1 / 2} \\
\sigma_{z}=\sqrt{\left\langle z^{2}\right\rangle}=\frac{1}{\left(2 \epsilon_{\|}\right)^{1 / 2}}\left(\frac{k_{\mathrm{B}} T}{U_{0}}\right)^{1 / 2}
\end{gathered}
\end{equation}
where $k_B$ is the Boltzmann constant.

 \begin{figure}[!t]
    \centering
    \includegraphics[width=0.45\textwidth]{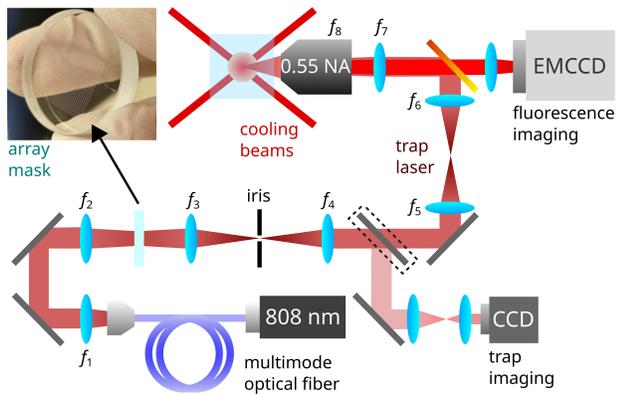}
    \caption{Schematic of the experimental apparatus for trapping laser-cooled cesium atoms in a dark trap array made with a broadband, spatially multimode 808 nm laser. Atoms are loaded into a magneto-optical trap (MOT), then transferred into a blue-detuned dark trap array which overlaps the MOT. The trap array pattern is formed using a passive optical mask and $4f$ filtering scheme, and the trap is imaged onto the atoms using a 0.55 NA custom objective. The dashed outline denotes a removable mirror which is used for imaging the trap at low power onto a CCD camera. An EMCCD camera captures 852 nm fluorescence from the atoms, which is collected through the same objective lens used for focusing the trap light and picked off with a long-pass dichroic. The labeled lenses have focal lengths $f_{1-8} = (4.5,500,250,75,200,30,150,22.8)$ mm.}
    \label{fig:schem}
\end{figure}

For a Gaussian beam with waist parameter $w_0$ we have $\epsilon_{\perp}=2 / w_0^{2}, \epsilon_{\|}=\lambda^{2}/(\pi^{2} w_0^{4}) = 1/z_R^2$. The bright aG has similar atom distributions but with numerical corrections close to unity. The results for Gaussian, aG, and dark aG potentials are summarized in Table \ref{tab:params}.

For the dark aG trap, the potential is dominated by a quartic term to lowest order in the radial direction. Using the Virial theorem again, we obtain 
\begin{equation}
    \begin{gathered}
    \langle \rho^4 \rangle = \frac{2}{3}\frac{k_B T}{U_0 \epsilon_{\perp}}\\
    \sigma_{\rho} = \left(\frac{2}{3}\frac{k_B T}{U_0 \epsilon_{\perp}} \right)^{1/4}
    \end{gathered}
\end{equation}
where $\epsilon_{\perp}$ is the coefficient of the quartic term in the radial expansion of the potential.

In practice, the radial and axial confinement provided by a trap potential is found by a parametric heating experiment to deduce the vibrational frequencies of the trapped atoms. These are well defined along directions for which the potential is harmonic to lowest order. For a trap which closely approximates a Gaussian potential of waist $w_0$, with trap depth $U_0$ for an atom of mass $m$, we have 
\begin{equation}\label{eq:harmonic_freqs}
    \begin{gathered}
    \omega_{\rho} = \frac{2}{w_0}\sqrt{\frac{U_0}{m}}\\
    \omega_z = \frac{1}{h z_R}\sqrt{\frac{2 U_0}{m}}
    \end{gathered}
\end{equation}
For a perfect Gaussian beam, $h=1$. For the bright aG beam, which diverges somewhat slower in the axial direction, $h=1.307$, and for the dark aG trap $h=0.997$. In the case of the dark aG trap, the radial profile is dominated by a quartic term, so the radial motion of trapped atoms will obey the unforced Duffing equation \cite{Kovacic2011}. Hence, a particular vibrational frequency is not well-defined in this case. A summary of the trap frequencies and spatial confinement is given in Table \ref{tab:params}.

\begin{figure*}[!t]
    \centering
    \includegraphics[width=1\textwidth]{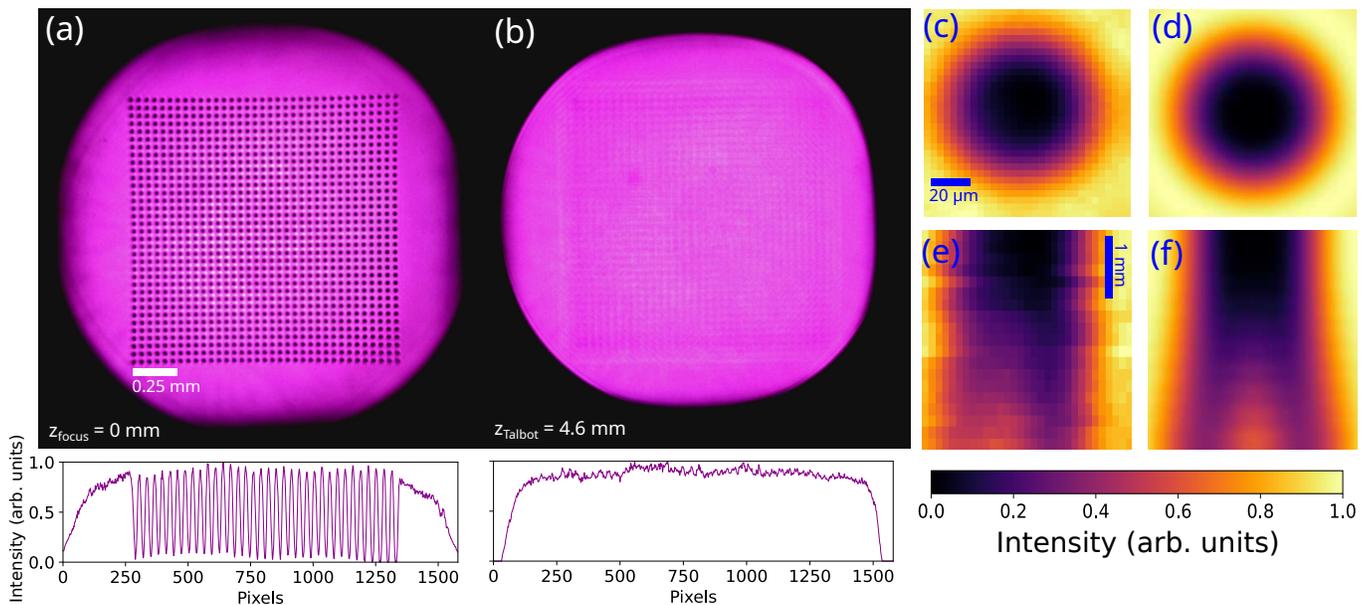}
    \caption{Dark trap intensity profile with broadband, spatially multimode light. \textbf{(a)} The focal plane of traps imaged onto a CCD, formed with a $4f$ filtering system with a magnification of 1/10, giving an array period of $43~\mu$m in the imaging plane. For the $\lambda= 805~ $nm light used, \textbf{(b)} the Talbot plane is located at $z=4.6$ mm, where the re-imaged traps have been strongly suppressed. \textbf{(c)} and \textbf{(e)} show a single trap site with 808 nm multimode light projected on the focal plane and along the propagation plane intersecting $\rho=0$. For qualitative comparison, \textbf{(d)} and \textbf{(f)} show numerical simulations of the trap using coherent single mode light}
    \label{fig:trap_intensity}
\end{figure*}

\section{Experiment} \label{sec:exp}

We demonstrate the proposed method of creating a blue-detuned dark trap array with up to 1225 sites for trapping single Cs atoms, and compare the use of an incoherent laser versus a coherent laser. The trap array is formed using a commercially fabricated (by LaserOptik) custom mask consisting of a $35 \times 35$ grid of partially transmitting disks of radius  $a=100~\mu$m ($|t_a|^2 = 0.49$; see Appendix \ref{sub:profiles}) and spatial period $d=430~\mu \textrm{m}$, positioned on a 25.4 mm diameter glass blank AR coated for a design wavelength of 825 nm. The experimental setup, shown in Fig. \ref{fig:schem}, is the same for trapping with both coherent and incoherent lasers except for the lasers themselves and the optics that precede the array mask. The array mask and $4f$ telescope is followed by an intermediate pick off for imaging the trap intensity with a magnification of roughly $1/10$, and the array is imaged into the science chamber to have a spatial period of 3 $\mu$m at the atoms. We emphasize that for a given array mask, the period of the array at the atoms and the trap depth can be tuned by changing the magnification of the imaging system.

\subsection{Incoherent trap}\label{sub:incoherent}
\begin{figure*}[!t]
    \centering
    \includegraphics[width=\textwidth]{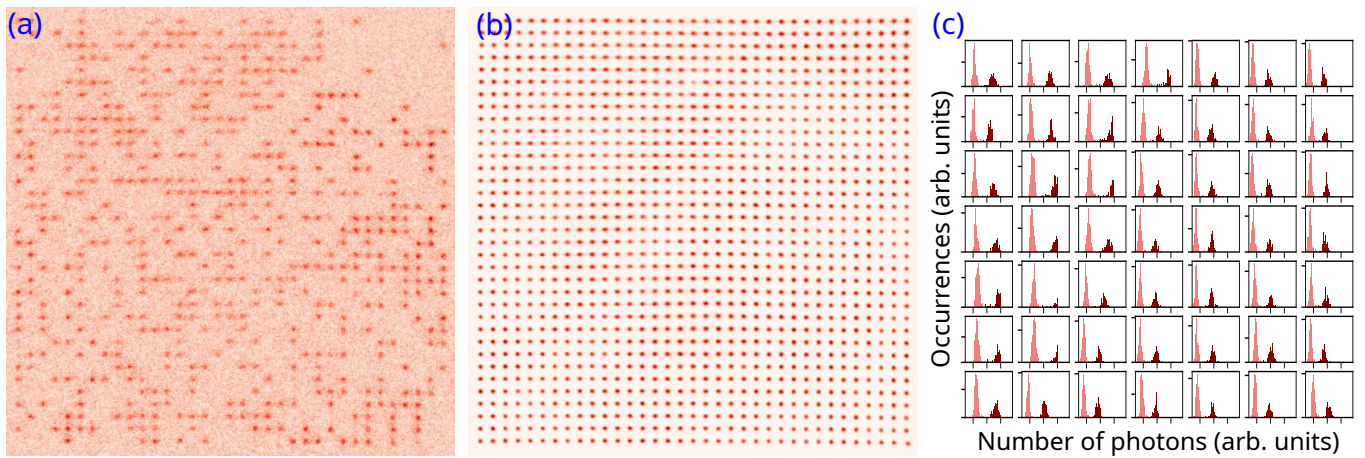}
    \caption{Single atom loading in a 1225-site array of dark traps formed with a broadband, spatially multimode laser. See section \ref{sec:exp}. \textbf{(a)} Single fluorescence image, showing stochastic loading of the trap array. The spacing between trap sites is about $3~\mathrm{ \mu m}$. \textbf{(b)} Averaged fluorescence image processed with independent component analysis. (see Appendix \ref{sub:imaging}). \textbf{(c)} Histograms of photocounts from regions of interests defined for the central 49 sites in (a),(b), showing the well-separated background (pink) and single atom (dark red) distributions.}
    \label{fig:bigarray}
\end{figure*}
First, we present single atom trapping in a 2D 1225-site  dark trap array with a high power spatially multimode, broadband 808 nm laser (Aerodiode CCMI, 35 W), which is blue-detuned with respect to  the Cs D-lines. Going forward, we will simply refer to this laser as the ``incoherent'' laser. 

The incoherent light is used in order to form the optical trap array without incurring the formation of Talbot plane traps, as shown in Fig. \ref{fig:trap_intensity}. For an explanation of the dominant mechanism in destroying the Talbot interference, see Appendix \ref{sec:talbot}. The entire array mask was illuminated and imaged onto a CCD following a $4f$ filtering setup. In order to have uniform trap depths across the array, it is crucial to uniformly illuminate the mask, which we do by re-imaging the core of a $200~\mu$m multimode fiber on to the mask. This is a good approximation to illuminating the mask with a top-hat intensity profile. An adjustable iris was used as the spatial filter in the $4f$ setup, and tuned to a radius of about 0.5 mm to minimize the trap center intensities as viewed on a CCD camera. To characterize the trap, we begin by stochastically loading single Cs atoms from a MOT into the array, for which we observe up to about 50$\%$ filling of the array. The trap light is switched on by controlling the laser current, instead of using an AOM, due to poor diffraction efficiency with the multimode light. Fluorescence images of atoms loaded into the array are shown in Fig. \ref{fig:bigarray}, with an average single atom loading rate in excess of 30$\%$ for the data shown. The loading rate  observed with coherent trapping light, discussed below, was about 50$\%$. We attribute the increase in loading rate with coherent light to the difference in relative intensity noise (RIN) between the two lasers (see Appendix \ref{sub:lifetime}). 

The axial and radial frequencies of the atoms in the trap were measured to be about 6 kHz and 44 kHz, respectively, by parametric heating from modulating the laser current with a sinusoidal source. For the particular mask used, we expect a trap which diverges faster in the axial direction compared to the expansion given in eq. (\ref{eq:Iag_dark2}), providing tighter axial confinement. For the radial direction, the trap closely matches a harmonic potential, and therefore the standard radial frequency relation (see eq. (\ref{eq:harmonic_freqs})) for a harmonic potential applies (see Appendix \ref{sub:profiles}). Hence there are three free parameters in the pair of trap frequency equations: the best-fit Gaussian waist $w_0$, trap depth $U_0$, and divergence parameter $h$. It is not straightforward to predict $h$ due to the unknown $M^2$ for the incoherent light, so we use the two trap frequencies to solve for $h$ and $U_0$, given a value of $w_0$ found using the known imaging magnification and a calibrated fit of the trap intensity. For a fit waist of 1.6 $\mu$m, the trap frequencies imply $U_0/k_B=462 ~\mu$K and $h=0.647$. The polarizability of the Cs $6\rm{S}_{1/2}$ states is $\alpha_0 = 0.66\times10^{-6}~\rm{\mu K/(W/m^2)}$ at 808 nm, and the estimated intensity at the atoms is $7.06 \times 10^8~\rm{W/m^2}$ (with about 20 W of light), giving an expected trap depth of $U_0/k_B=466 ~\rm{\mu K}$. This agrees with the value implied by the trap frequencies to within a few percent, and $h$ is of the same order as the numerical prediction with coherent light. 

The observed atom lifetime in the incoherent trap was about 40 ms, in comparison to about 5 s measured with the coherent trap to be discussed next. We attribute this difference to the RIN of each laser (see Appendix \ref{sub:lifetime}). Nevertheless, we emphasize there is nothing fundamentally prohibitive for obtaining reasonable trap lifetimes in  multimode optical traps \cite{WHung2015,Povilus2005}. Moreover, intensity-noise limited trap lifetimes can be significantly improved  using well developed noise reduction techniques such as the method used in \cite{YuWang2020}, which uses an AOM and EOM for slow and fast noise reduction, respectively. We note that the poor diffraction efficiency of an AOM with multimode light is not a barrier to intensity stabilization as it was to using an AOM as an optical switch. This is because using an AOM for intensity stabilization requires dumping a relatively small amount of power into the diffracted order. Furthermore, the intensity stabilization scheme in \cite{YuWang2020} could be modified to use two EOMs instead of one AOM and one EOM, if diffraction efficiency is a concern. 

\subsection{Coherent trap}\label{sub:coherent}
The same trap characterization experiments were repeated using a coherent 825 nm laser (Moglabs MSA003 tapered amplifier), where the only change to the experimental setup was the optics for collimating the trap light before the array mask. The mask was illuminated using a Gaussian beam which was collimated after being spatially filtered by a single mode fiber. Despite the Gaussian illumination of the mask, which leads to a varying trap depth across the array, we still observe traps which go completely or near completely dark across the array, indicating that uniform mask illumination is not  strictly required.

Again, we characterize the confinement by measuring the trap frequencies and trap lifetime. The trap depth for different sites varies according to the Gaussian intensity envelope across the array. We will therefore restrict our analysis to values near the center of the array, where the traps are deepest. The peak intensity at the atoms is given approximately by $I = 2P/\pi w_{\rm env}^2$, where $P=260$ mW is the laser power and $w_{\rm env}=12~\mathrm{ \mu m}$ is the Gaussian envelope waist at the atoms. This yields about 20 to 25 sites which are deep enough for trapping. To parametrically heat the atoms, the trap light intensity was modulated by varying the RF power to an AOM placed before the single mode fiber delivering light to the $4f$ filtering setup. We measure radial and axial frequencies of about 52.5 kHz and 7.2 kHz, respectively. The rate of axial divergence, which is faster than that of a Gaussian beam, has a divergence parameter $h=0.65$ predicted by a Fresnel diffraction simulation for the parameters of the mask used (see ). Assuming this value of $h$, the trap frequencies imply a trap depth of $1.4$ mK and waist $w_0=1.81~\mu$m. This waist is consistent with the value of $1.89~\mu$m deduced from fitting the trap in the intermediate imaging plane to within a few percent, and the trap depth is within our uncertainty in measuring the intensity. Lastly, the trap lifetime was found to be nearly 5 s, measured in the same site for which we report the trap frequencies.

\section{Conclusion}

We have demonstrated a simple technique for creating optical trap arrays using only passive optical components.  The design lends itself to being scalable, as a larger trap array can be created with a larger mask grid and more optical power. This is an important point for scaling up neutral atom qubit arrays, for which the limiting factors are optical power and the performance of the trap imaging system\cite{Morgado2021}. As the design is based on only passive components, it is free from noise associated with active spatial light modulators such as DMDs. 

The design presented is versatile, lending itself to several compelling variations. Firstly, the same working principle can be used with a different mask to simultaneously create bright and dark arrays for trapping two-species. This can be done using a single trapping wavelength, without requiring any additional experimental footprint compared to one-species traps. Secondly, alternative Fourier filters can be used to create traps which provide significantly higher localization of trapped atoms (see Appendix \ref{sec:tighter}).

We have also demonstrated the use of a spectrally and spatially multimode trap for mitigating the Talbot effect, which leads to unwanted out-of-focus traps. However, the secondary arrays of traps formed with coherent light allows for creating a three-dimensional trap array, which may be advantageous for multi-dimensional architectures\cite{YWang2016,Barredo2018}.   



\section*{Acknowledgments}
This material is based upon work supported by the
U.S. Department of Energy Office of Science National
Quantum Information Science Research Centers as well as support from NSF Award 2016136 for the QLCI center Hybrid
Quantum Architectures and Networks, NSF award 1806548,  and U.S. Department
of Energy, Office of Science, Office of High Energy Physics,
under Award No. DE-SC0019465.  


\bibliography{rydberg,optics,qc_refs,atomic,saffman_refs}

\appendix

\section{Robustness of $4f$ filtered traps}\label{sec:sensitivity}

A natural question to pose is to what extent the $4f$ filtering approach to generating traps is sensitive to parameters, such as the mask transmission amplitudes, Fourier filter radius, and the intensity profile used to illuminate the mask. Here we address this for the case of generating dark traps. The metric we find most useful is simply to quantify the relative intensity at the centers of the traps, that is, $I_2(0,0)/I_0$ (eq. (\ref{eq:Iag_dark})). The derivation of the dark trap found the condition for completely dark traps given zero relative phase between the mask aperture and background transmission amplitudes, $t_a,t_b$, and a choice of filter radius $b$ corresponding to blocking the Airy disk beyond the first minima. 
\begin{figure}[!ht]
    \centering
    \includegraphics[width=0.5\textwidth]{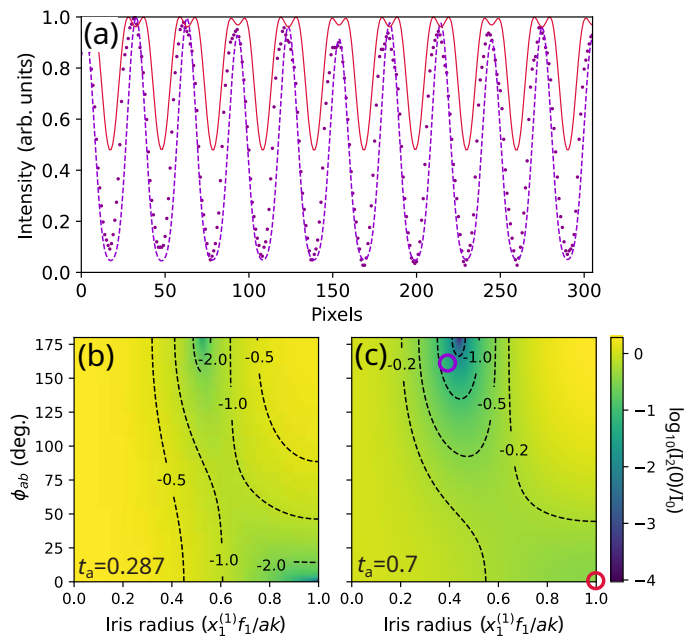}
    \caption{Numerical analysis of dark aG trap profile dependence on relative mask phase $\phi_{ab}$ and Fourier filter (iris) radius. The iris radii here are given as dimensionless by dividing by the value used in the derivation in the main text, $b=x^{(1)}_1 f_1/a k$. \textbf{(a)} Trap profile data (purple points) from Fig. \ref{fig:trap_intensity}a compared to two Fresnel diffraction calculations: one with parameters which give a similar profile (purple dashes), and one with parameters corresponding to the expected trap given the specified parameters of the mask used (red line). Both parameter sets  use $f_1=500$mm,$f_2=50$mm, $\lambda=805$nm, with the former using ($\phi_{ab}$, iris radius) = ($160^{\circ}$,0.4), and ($0^{\circ}$,1) for the latter. \textbf{(b)} and \textbf{(c)} show the dark trap center intensity (log scale) versus $\phi_{ab}$ and iris radius. The choices of $\phi_{ab}$ and iris radius used to generate the numerical curves in (a) are marked with rings of the same color in (c).}
    \label{fig:idark_vs_phi_b}
\end{figure}

However, the trap can be made near completely dark for other choices of these parameters. Fig. \ref{fig:idark_vs_phi_b} shows the trap center intensity for two values of $t_a$ (and $t_b=1$), as a colormap versus the relative phase $\phi_{ab}$ between $t_b,t_a$ and the Fourier filter or iris radius cast in dimensionless units. For $t_a=0.287$, as derived as optimal above, we see the trap remains dark to within about $10 \%$ for up to around 20 degrees of relative mask phase, indicating the robustness of the design with respect to deviance in parameters from manufacturing of a real mask.

\section{Tighter traps with alternative Fourier plane masks}\label{sec:tighter}
\begin{figure}[!ht]
    \centering
    \includegraphics[width=0.5\textwidth]{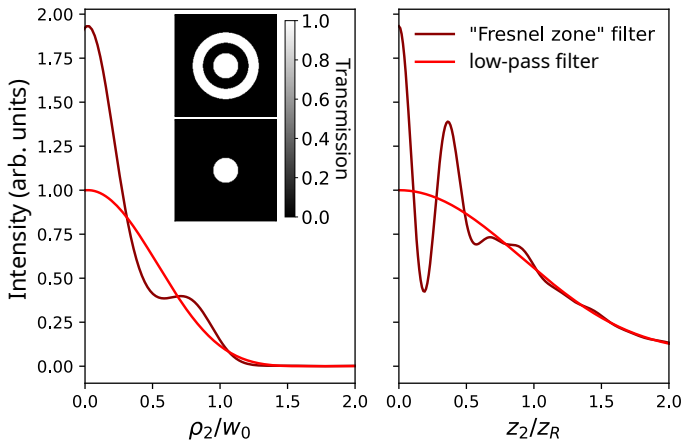}
    \caption{Comparison between an aG beam produced with a low-pass Fourier filter (see main text) and a modified version using a "Fresnel zone" Fourier filter. The horizontal axes are scaled to the best fit Gaussian waist $w_0$ and corresponding Rayleigh range $z_R=\pi w_0^2/\lambda$ for the aG beam. The two Fourier filter variants are shown in the inset, and the radii of the transmitting rings and central aperture are given in the text.}
    \label{fig:zonefilter}
\end{figure}

It is possible to create tighter trap profiles than those presented in the main text, by replacing the simple Fourier plane iris with more complicated amplitude or phase masks. For example, a higher efficiency (eq. (\ref{eq:brighteff})) bright trap can be created by using a modified Fourier plane amplitude mask which transmits certain higher spatial frequency bands. This amounts to adding spatial notch filters, i.e. transmitting rings, to a low-pass filter as used in the main text. It can be shown that the transparent rings should have inner and outer radii equal to $f_1 x_{2n}^{(1)}/(ak)$, $f_1 x_{2n+1}^{(1)}/(ak)$ respectively, for the $n^{th}$ ring from the center, for $n>1$. The central aperture has radius $f_1 x_{1}^{(1)}/(ak)$. This type of filter, which we refer to as a Fresnel zone filter, is shown with  only one ring in Fig. \ref{fig:zonefilter} alongside the resulting trap profiles compared to those of a standard aG beam. It is apparent from the figure that the resulting traps have much stronger confinement than is obtained with the simple low pass filter. Compared to the bright aG trap, the trap generated with the zone filter has roughly 1.9 times higher efficiency (corresponding directly to the trap depth), with about 2.2 and 10 times tighter confinement in $\rho$ and $z$, respectively. The zone filter does introduce weak secondary traps away from the origin. Atom trapping in the secondary traps can be avoided, either by slowly increasing the trap power during the loading phase, or by using atom rearrangement techniques to place atoms in the central trap. 

\section{Talbot Trap Mitigation}\label{sec:talbot}

\begin{figure}[!th]
    \centering
    \includegraphics[width=0.5\textwidth]{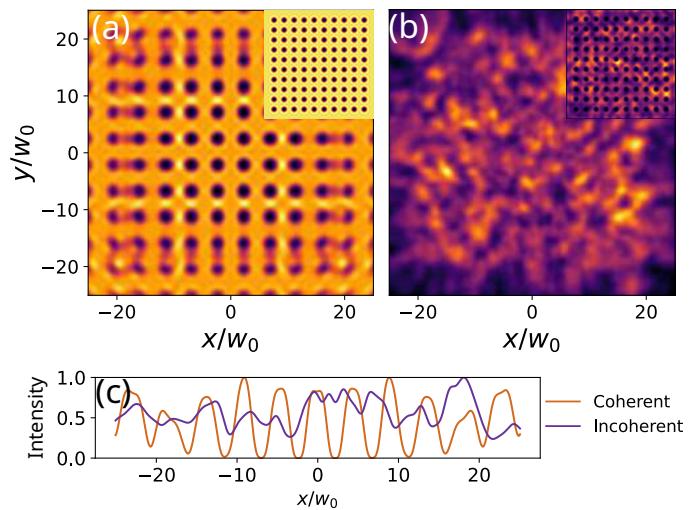}
    \caption{Comparison of simulated Talbot plane trap formation for a 10 by 10 dark aG array formed with \textbf{(a)} spatially coherent monochromatic light and \textbf{(b)} spectrally and spatially incoherent field. For the incoherent light, 21 frequency components spanning $\Delta \lambda_{\rm FWHM} = 3$ nm, each with 200 Hermite-Gaussian modes with random phases. The insets show the the focal plane intensity. The plot axes are scaled to the best fit Gaussian waist, $w_0=0.943 a(f_2/f_1)$. \textbf{(c)} Line profile comparison of a row of traps from (a) and (b), showing the washed out Talbot interference for the multimode light. For both simulations, $\lambda_0=825$ nm, $d=430~\mu$m, and $a=100~\mu$m, and a magnification of 1/100.}
    \label{fig:talbotcompare}
\end{figure}

In cold atom experiments using periodic optical trap arrays, it is a known problem that secondary traps, which form due to the Talbot effect, lead to out-of-focus trapped atoms. This contributes additional background in fluorescence measurements used for atom state detection and atom rearrangement \cite{graham2019}. The Talbot effect is a well-known phenomenon in optics, in which a field which is spatially periodic in the transverse plane will be reimaged after propagating a distance equal to 
\begin{equation*}
    z_{\text{Talbot}} = \frac{2d^2}{\lambda}
\end{equation*}
where $d$ is the transverse spatial period of the field and $\lambda$ is the wavelength of the field. Numerical simulations of the first Talbot planes for dark and bright trap arrays are shown in Fig. \ref{fig:talbotcompare} We propose a solution using a spatially and spectrally incoherent laser field. 

The Talbot effect can be mitigated by use of a spatially multimode laser, which has a limited spatial coherence. In what follows we will limit the numerical analysis to the case of a dark trap array, although much of the discussion is equally applicable to bright traps. In addition to being spatially multimode, it is desirable to have many frequency components as well. This ensures that the bright region of the trap does not exhibit a speckle pattern, which could create additional unwanted traps. This section will outline results of numerical simulation. The experimental confirmation of these predictions shown in Sec. \ref{sec:exp} of the main text. 

The models presented here use Fresnel diffraction theory, which is readily computed using fast Fourier transform (FFT) algorithms that are available in modern programming languages \cite{kelly2014}. This modeling can be done efficiently on a typical laptop. For computational efficiency, the results presented below are for 10-by-10 trap arrays. All of the Talbot plane intensity plots shown are normalized to the respective peak focal plane intensity, to provide a clear sense of the relative trap depth. For the simulations discussed, the laser wavelength is $\lambda =825 $ nm, mask grid period $d=430~ \mu$m, mask spot radius $a=100~\mu$m, aperture transmission $t_a=0.287$. However, the results are presented in a way that is independent of the choice of parameters such as focal lengths and wavelength.

Let us first consider an array of optical traps formed by a laser which has many frequency components, but is spatially coherent. Far-off resonance optical traps do not require narrow-linewidth lasers, and hence the use of a free-running, broadband laser is feasible. Naively, we may assume that if the coherence length of the laser is less than the Talbot length of the trap array, the Talbot traps will be suppressed. To quantify how short the coherence length must be in order for the Talbot traps to be washed out, we need to consider 1) the axial displacement of the Talbot plane for a given $\Delta \lambda$ from the central trap wavelength $\lambda_0$, and 2) the axial depth of the Talbot traps. The depth of the Talbot plane is $\delta Z_{\textrm{Talbot}} \sim \pi w_0^2/\lambda$ where $w_0=0.947af_2/f_1$ is the waist of the trap (for the case of the bright array), and the amount that the Talbot plane is displaced for a change in $\lambda$ is $2 \Delta \lambda d^2/\lambda^2$. Combining these, we find that in order to displace the Talbot plane by half of its axial depth, we require a laser with a spectral full-width of 
\begin{equation}
    \Delta \lambda_{1/2} = \frac{\pi \lambda}{2} \frac{a^2}{d^2} \frac{f_2^2}{f_1^2}
\end{equation}
For $a=100~\mu$m and a magnification of $1/100$, a laser with a FWHM of more than 10 nm is required to sufficiently wash out the Talbot plane. Typically, even lower quality laser diodes have linewidths of at most a few nm, insufficient for mitigating Talbot trap formation for the trap parameters considered. However, as we will show, spectral incoherence still plays an important role in the demonstrated solution to the Talbot problem.

We find that a practical and cost-effective approach to mitigating Talbot traps is using a laser which is both spectrally and spatially incoherent, which are readily available at high power. By using a spatially  multimode field, there is no longer spatial coherence which is required for forming the Talbot planes. A monochromatic but spatially incoherent field, as can be made by passing a laser with a single spatial mode through a  multimode fiber, has a speckled intensity and hence is not a good solution for creating trap arrays. A spectrally and spatially  multimode field will, however, not be speckled. This is explained by thinking of each spectral component as having its own associated speckle pattern. These speckle patterns add incoherently, i.e. the intensities rather than the fields are summed, resulting in a more uniform intensity pattern.

We model spatially incoherent fields by summing Hermite-Gaussian fields $A_{i,j}$ from $A_{0,0}$ to some highest-order $A_{m,n}$, with each spatial mode being given a random phase pulled from a flat distribution.  This process is done for each spectral component of the field and scaled by $\sqrt{L(\nu, \nu_0)}$ where $L$ is the Lorentzian function describing the spectral width of the laser. Each speckle field is then propagated through the $4f$ array generator and the intensities at the output are summed, yielding the output shown in Fig. \ref{fig:talbotcompare}b.

\section{Experimental details}\label{sec:experiment}

\subsection{Array mask and observed trap profiles}\label{sub:profiles}
The experimental demonstration of the proposed trap design used a commercially fabricated array mask consisting of partially transmitting disks (what we have previously been calling ``apertures'') set on an AR-coated 1'' diameter glass blank. The requested transmission of the disks was $|t_a|^2=0.49\pm 0.04$, with a relative phase of $0 \pm 10 $ deg. between the disks and AR coated background, designed for $\lambda = 825$ nm. The disks have radius $a=100~\mu$m and arranged in a 2D square grid with spatial period $d=430~\mu$m. The expected trap profile for these mask parameters does yields traps which are only about 50$\%$ dark (Fig. \ref{fig:idark_vs_phi_b}a). However, given the observed nearly dark traps and numerical analysis, we infer that the relative phase imparted on the light between the fully and partially transmitting regions must be outside of the specified tolerance in order to explain the observed trap profiles. 

\begin{figure}[t!]
    \centering
    \includegraphics[width=0.45\textwidth]{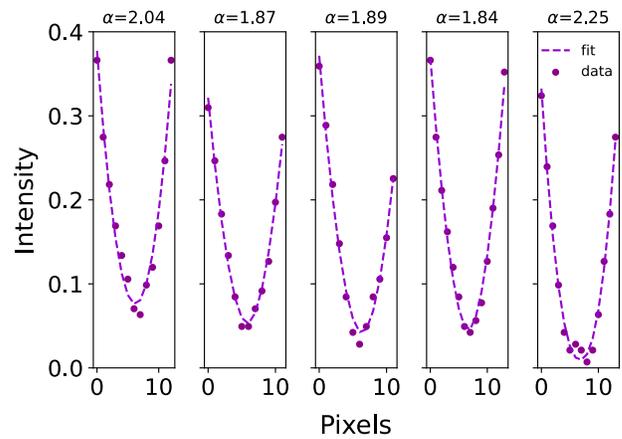}
    \caption{A row of adjacent trap site profiles from the data shown in Fig. 2, where each site has been fit to a form $a\rho^\alpha+b$. The data was clipped above the $1/e$ intensity to ensure fitting to the lower order portions of the traps. It is clear that the experimentally observed radial profiles are nearly quadratic, rather than the quartic profiles expected for $t_a=0.287t_b$ with zero relative mask phase.}
    \label{fig:site_fits}
\end{figure}

Because a direct phase measurement of the array mask is difficult, we present numerical analysis to explain the observed trap profile. These results are summarized in Fig. \ref{fig:idark_vs_phi_b}. The observed traps can be made dark to within $10\%$ of the peak intensity by adjusting the iris radius. This constrains the possible values of $\phi_{ab}$ and iris radius that are consistent with the trap we observe. For example, a Fresnel diffraction calculation for $t_a=0.7$, iris radius $b=0.4x^{(1)}_1 f_1/a k$ and $\phi=160^{\circ}$, gives a result that agrees well the observed trap profiles (Fig. \ref{fig:idark_vs_phi_b}a, purple dashes). This choice of parameters is marked with a purple ring in Fig. \ref{fig:idark_vs_phi_b}b.

Moreover, the axial divergence rate of the trap is in good agreement with the measured trap frequencies. It can be shown that the quadratic term of the axial expansion for $\phi_{ab}=160^{\circ}$ and $b=0.4x^{(1)}_1 f_1/a k$ is equal to $2.356(z/z_R)^2$, giving a divergence parameter $h=1/\sqrt{2.356}=0.65$. This is an additional advantage conferred by the mask parameters, as the axial confinement is $\sim 30\%$ tighter compared to the expansion rate using $t_a=0.287$ and $\phi_{ab}=0$. As stated in the main text, the measured trap frequencies predict a trap waist and depth consistent with estimates of this value of $h$ is assumed in eq. (\ref{eq:harmonic_freqs}).

Lastly, the measurement of well-defined vibrational frequencies discussed in the main text is a result of the quadratic profile of the observed traps. This is shown with power fits to a row of adjacent sites in Fig. \ref{fig:site_fits}. It can be shown that the quadratic coefficient in the radial expansion of the trap profile, for $t_a=0.7$, is only finite and positive for certain values of the relative mask phase $\phi_{ab}$ and iris radius, including $\phi_{ab}=160^{\circ}$ and $b=0.4x^{(1)}_1 f_1/a k$.

\subsection{Atom lifetime in incoherent trap}\label{sub:lifetime}
\begin{figure}[t!]
    \centering
    \includegraphics[width=0.45\textwidth]{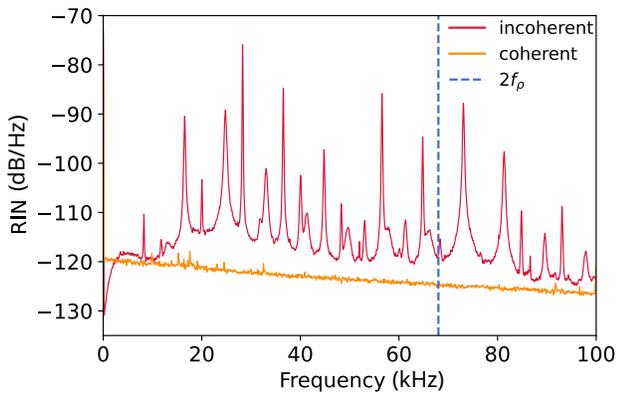}
    \caption{RIN spectrum for the incoherent 808 nm laser and coherent 825 nm lasers used for atom trapping. The  atom lifetime in the incoherent trap was measured to be 40 ms, in contrast to a lifetime of at least 5 s measured with the coherent trap. The dashed blue line corresponds to the twice the measured radial vibration frequency, where the 808 nm RIN level implies a lifetime in excess of 8 s. However, the many large frequency spikes can induce off-resonant heating.}
    \label{fig:rin}
\end{figure}

We attribute the observed characteristic lifetime of atoms in the incoherent trap, about 40 ms, to the RIN measured on the laser, shown in Fig. \ref{fig:rin}. While the level of the RIN near 68 kHz (twice the measured observed radial resonance) implies a lifetime of over 8 seconds \cite{Gehm1998}, there is a lower frequency bump nearby and several high amplitude spikes. Because the atom is heated by these other components as well, we take this noise to be the explanation of the observed poor lifetimes. 
The RIN for a coherent laser used for trapping is shown for comparison, for which we measured a lifetime around 5 s. Note that the seed laser for the coherent tapered amplifier used for experiments described in the main text (for which a 7 s trap lifetime was observed) died before RIN data could be measured. 

\subsection{Fluorescence imaging}\label{sub:imaging}
The fluorescence images of trapped atoms were captured on a  Hamamatsu C9100-13 EMCCD with 100 ms exposure time. The imaging light was 852 nm, red detuned from the D2 line cooling transition $F=4 \leftrightarrow F'=5$ by about 9 times the natural linewidth, with an intensity around three times the saturation intensity. The single shot image of trapped atoms in Fig. 2a is divided by the average background for the stack of 300 shots taken for that data to account for an uneven intensity pattern on the shots due to a gain variation on the EMCCD sensor itself. The large image was processed with independent component analysis, which has the effect of de-blurring the trapped atoms. However, due to the size of the image (512 by 512 pixels), this process was only effective when the image was broken into smaller chunks. Ultimately, 25 sub-regions in the image stack were processed, and the the summed projected images were stitched back together. The result exhibited a uniform background in each sub-image, with a slight variation between the background level in adjacent sub-regions. Hence, for visual purposes only, the resulting image was plotted with the pixels valued less than 20$\%$ of the maximum intensity clipped.

\end{document}